\documentstyle[11pt,epsf]{article}

\def\beq{\begin{equation}}
\def\eeq{\end{equation}}
\def\eq{\end{equation}}
\def\ba{\begin{eqnarray}}
\def\ea{\end{eqnarray}}

\def\centeron#1#2{{\setbox0=\hbox{#1}\setbox1=\hbox{#2}\ifdim
\wd1>\wd0\kern.5\wd1\kern-.5\wd0\fi
\copy0\kern-.5\wd0\kern-.5\wd1\copy1\ifdim\wd0>\wd1
\kern.5\wd0\kern-.5\wd1\fi}}
\def\ltap{\;\centeron{\raise.35ex\hbox{$<$}}{\lower.65ex\hbox{$\sim$}}\;}
\def\gtap{\;\centeron{\raise.35ex\hbox{$>$}}{\lower.65ex\hbox{$\sim$}}\;}

\def\singleandthirdspaced{\baselineskip=\normalbaselineskip\multiply
    \baselineskip by 130\divide\baselineskip by 100}

\def\dslash{\not{\hbox{\kern-2pt $\partial$}}}
\def\Dslash{\not{\hbox{\kern-4pt $D$}}}
\def\Oslash{\not{\hbox{\kern-4pt $O$}}}
\def\Qslash{\not{\hbox{\kern-4pt $Q$}}}
\def\pslash{\not{\hbox{\kern-2.3pt $p$}}}
\def\kslash{\not{\hbox{\kern-2.3pt $k$}}}
\def\qslash{\not{\hbox{\kern-2.3pt $q$}}}
\def\epsilonslash{\not{\hbox{\kern-2.3pt $\epsilon$}}}
\newcommand{\newc}{\newcommand}
\newc{\qbar}{{\overline q}}
\newc{\Kahler}{K\"ahler }
\newc{\deltaGS}{\delta_{\rm GS}}
\begin{document}
\begin{titlepage}
\begin{flushright}
{\large hep-th/0111235 \\ SCIPP-01/12 \\ SU-ITP-01/21 \\ NSF-ITP-01-58 \\}

\end{flushright}

\vskip 1.2cm

\begin{center}

{\LARGE\bf
Brane World Susy Breaking
}

\vskip 1.4cm

{\large Alexey Anisimov$^1$, Michael Dine$^1$, Michael Graesser$^1$, and
 Scott Thomas$^{2,3}$}
\\
\vskip 0.4cm
%{\it $^a$Stanford Linear Accelerator Center,
%     Stanford CA 94309} \\ \vskip 1pt
{\it $^1$Santa Cruz Institute for Particle Physics,
     Santa Cruz CA 95064  } \\ \vskip 1pt
{\it $^2$Physics Department, Stanford University, Stanford, CA 94305  } \\
 \vskip 1pt
{\it $^3$Institute for Theoretical Physics, University of California,
Santa Barbara, CA 93106}\\

%{\it $^c$Physics Department,
%     University of California,
%     Santa Cruz CA 95064  }

%\vskip 0.6cm

%{\large Scott Thomas}
%\\
%\vskip 0.4cm
%{\it $^a$Stanford Linear Accelerator Center,
%     Stanford CA 94309} \\
%{\it $^c$Physics Department,
%     University of California,
%     Santa Cruz CA 95064  }

\vskip 4pt

\vskip 1.5cm

\begin{abstract}

In brane world models of nature, supersymmetry breaking is often
isolated on a distant brane in a higher dimensional space.
The form of the Kahler potential in generic string and M-theory    
brane world backgrounds is shown to give rise to tree-level
non-universal squark and slepton masses.  This results from
the exchange of bulk supergravity fields and warping of the
internal geometry.  This is contrary to the notion that bulk locality
gives rise to a sequestered no-scale form of the Kahler potential
with vanishing tree-level masses and solves the supersymmetric flavor
problem.  As a result, a radiatively generated anomaly mediated
superpartner spectrum is not a generic outcome of these theories.

\end{abstract}

\end{center}

\vskip 1.0 cm

\end{titlepage}
\setcounter{footnote}{0} \setcounter{page}{2}
\setcounter{section}{0} \setcounter{subsection}{0}
\setcounter{subsubsection}{0}

%%%%%%%%%%%%%%%%%%%%%%%%%%%%%%%%%%%%%%%%%%%
%%%%%%%%%%%%%%%%%%%%%%%%%%%%
\singleandthirdspaced

%\begin{document}

\section{Introduction}
\label{sec:intro}

In many pictures for the origin of supersymmetry breaking,
gravity and
non-renormalizable operators, play a crucial role.
In some, higher dimensional space-times play an important
role.
The only framework in which
such questions can at present be consistently
addressed is string or M-theory.
To actually show that string or M-theory
predicts low energy supersymmetry, much less
a particular form for the pattern of soft breakings,
is beyond present capabilities.  However, given the assumption
of an approximate low energy supersymmetry, one can survey supersymmetric
states of string or M-theory and look for generic features.
%  We might hope to address
%these questions by someday finding a solution of the theory,
%and determining its features.  But this is unlikely to
%happen any time soon.
%However,
%these sorts of questions can be addressed
%by looking for generic features of string or M-theory
%ground states.
%Particular mechanisms both for supersymmetry breaking, and
%the mechanism of mediation to the visible sector, may be postulated
%and the consequences examined.

%It is, however, possible even within our current state
%of knowledge, to assess the plausibility of some
%proposed supersymmetry breaking/mediation mechanisms.  Some
%proposed mechanisms presume certain generic features of
%a fundamental theory, and these can be studied by examining broad
%classes of string states.  In other words, we can sometimes hope
%to argue that string theory {\it does not} predict some particular
%phenomenon.

Among proposed mediation mechanisms, one which can be studied
along these lines 
%with string or M-theory 
is Brane World Supersymmetry Breaking (BWSB).
%(this class of proposals includes ``Anomaly Mediation"
%\cite{randallsundrum,lutyetal}
%and ``Gaugino Mediation"\cite{gauginoa,gauginob}).
The defining feature of this picture is that
visible sector fields are localized on a brane while supersymmetry
breaking is isolated on a physically separated hidden sector brane,
all within a compact higher dimensional space.
Such a structure is a plausible outcome of string or M-theory.
One might expect that bulk locality would have striking consequences
for such ground states, and indeed it has been argued that
the Kahler potential takes a particular no-scale or sequestered form.
This leads to vanishing tree-level scalar masses.  The leading
contributions to scalar masses have been argued to be due to
anomaly mediation, yielding a solution to the supersymmetric
flavor problem \cite{randallsundrum,lutyetal}.

In this note we determine the four--dimensional Kahler potential which
couples the visible and hidden sector branes in a number
of BWSB backgrounds.  In all of the cases we can analyze,
the Kahler potential is not of the sequestered no-scale form.
With supersymmetry breaking isolated on the
hidden sector brane, these Kahler potentials generally give
rise to tree-level soft scalar masses for visible sector
squark and slepton fields.
The masses are of order the four--dimensional gravitino
mass and are generally not universal.
The leading brane--brane couplings contained within the Kahler
potential which give rise to the tree-level masses
can be understood in these examples as arising from
exchange of bulk supergravity fields.
Additional corrections to the tree-level masses arise
from warping of the higher dimensional compact space.
Without additional assumptions about flavor,
the non-universal scalar mass matrices are not necessarily aligned
with the quark and lepton mass matrices, and dangerous
sflavor violation can occur.
So physically separating the visible and hidden sector branes within
a higher dimensional space is not alone enough to give
a predictive (anomaly mediated) spectrum and solve
the supersymmetric flavor problem.

The rest of this paper is organized as follows.  After reviewing
BWSB in section 2,
general macroscopic considerations based on extended supersymmetry
are employed in section 3 to determine the leading form of the
Kahler potential in a number of BWSB backgrounds, including
Type I$^{\prime}$ theory, Type IIB theory and heterotic M-theory.
With hidden sector supersymmetry breaking
these Kahler potentials give rise to non-universal tree-level scalar
masses which are the same order as the four--dimensional
gravitino mass.  We then determine the microscopic origin of
these apparently non-local
brane--brane interactions; they arise from the exchange
of bulk fields.
%The origin of various apparently non-local terms in both
%types of theories are explained
We note in section 4 that there are further corrections
to the lowest order form of the Kahler potential and therefore
soft scalar masses in both types of theories.  These can be thought
of as arising from warping of the internal bulk geometry by the brane.
In the Horava-Witten theory these
are not likely to be particularly small.
%Similar types of corrections are exhibited in the Type I
%theory.
We also speculate on special circumstances under which
the no-scale form of the Kahler potential and vanishing
tree-level scalar masses may arise, but argue that these
are not likely to be generic.
More examples of BWSB backgrounds (including microscopic
descriptions and leading corrections in various limits),
the inclusion of gaugino masses, and a discussion of the
closely related mechanism of gaugino mediation \cite{gauginoa,gauginob},
are presented elsewhere \cite{longpaper}.

%It should not be particulary suprising that such terms in fact arise
%since there is no sense in which the branes are far apart
%in the low energy four--dimensional theory.

% ----------------------------------------------------------

\section{The Brane World Picture}

In order to illustrate
the coupling between visible and hidden sector branes it is
convenient to work in the conformal or supergravity frame.
%The authors of \cite{randallsundrum} begin by writing
The bosonic
part of the four--dimensional supergravity action in this frame
is \cite{cremmer}
\ba
{\cal L}& = & {f \over 6} {\cal R}_{4}
-f_{i \bar{j}} \partial_{\mu} \varphi_i \partial_{\mu}
\varphi^* _{\bar{j}} - {1 \over 4 f}
(f_i \partial_{\mu} \varphi_i -\hbox{h.c.})^2 + \cdots
\nonumber \\
& & + f_{i \bar{j}} F_i F^* _{\bar{j}} +
\vert F_{\Phi} \vert ^2 f
+(W_i F_i + f _i F_{\Phi}^{*} F_i  + 3 F_{\Phi} W +\hbox{h.c.})
\label{cremmerlag}
\ea
where
$f$ is the field dependent supergravity function which multiplies
the four--dimensional Einstein term, and $W$ is the superpotential.
The supergravity $f$ function and Einstein frame Kahler
potential are related by
\beq
%f=-3 e^{-K/3} ~.
K= - 3 \ln ( -f/3) ~.
\eeq

In supergravity frame, the couplings of interest
between the branes reside in the $f$ function.
%The starting point of the sequestered hypothesis is the assumption
%that for fields on well-separated branes, the function $f$
%separates as:
A special class of $f$ functions is the separable form
\beq
f(T_i,Q_i,\Sigma_i) =
f_{\rm vis}(Q_i)+  f_{\rm hid}(\Sigma_i) - f_{\rm mod}(T_i+T_i^{\dagger})
\label{fsep}
\eeq
where $Q_i$ and $\Sigma_i$ are visible and hidden sector fields
respectively and $T_i$ are moduli.
In this case non-derivative couplings between fields on
the different branes vanish since $f_{i \bar j} =0$.
The separable form (\ref{fsep}) has been referred to as
sequestered and argued to arise in BWSB
backgrounds \cite{randallsundrum}.
This may seem plausible given that
bulk locality might lead one to expect some sort of
decoupling of fields on the physically separated
visible and hidden sector branes.
%In  particular, one might imagine that there is a natural frame in
%which the lagrangian breaks up into this form.
%After all, on the
%branes, one might expect that the kinetic terms for the scalar
%fields are canonical, and in any case would not involve couplings
%of the fields on the different branes.
The Kahler potential associated with the separable form
(\ref{fsep}) is of the no-scale type.
With canonical tree-level kinetic terms for the visible
and hidden sector fields,
$f_{\rm vis}=3 \hbox{tr}Q^{\dagger}_i Q_i$ and
$f_{\rm hid}=3 \hbox{tr}\Sigma^{\dagger}_i \Sigma_i$,
and only a single modulus $T$, the no-scale Kahler potential is
\beq
K = -3 \ln (
  f_{\rm mod}(T+T^{\dagger}) - \hbox{tr}Q^{\dagger}_i Q_i
  -  \hbox{tr}\Sigma^{\dagger}_i \Sigma_i )  ~.
\label{sequesteredkahler}
\eeq
A non-vanishing auxiliary component for either a hidden sector
field, $F_{\Sigma _i}$, or modulus, $F_{T}$, does
not give rise to visible sector scalar masses.
In Einstein frame, this seems to be the result of a miraculous
cancellation, depending on the logarithmic form and the
prefactor 3.
In the supergravity frame, however, it is a result of the separable
form of the $f$ function.

With a no-scale sequestered Kahler potential (\ref{sequesteredkahler})
the leading contributions to
scalar and gaugino masses arise from anomalous effects
(non-anomalous one- and two-loop contributions with bulk
scalar exchanges have been argued to be highly 
suppressed \cite{randallsundrum}).
The one-loop gaugino and two-loop scalar
anomaly mediated contributions to masses are given by
\beq
m_g = -b_0 {g^2 \over 16 \pi^2} m_{3/2}~~~~~~~
\tilde m_q^2 = {1 \over 2} c_0 b_0 \left( {g^2 \over 16 \pi^2} \right)^2
  \vert m_{3/2} \vert^2
  \label{anomalymasses}
\eeq
where $b_0$ and $c_0$ are the leading beta function and
anomalous dimension coefficients respectively (for vanishing
Yukawa couplings), and $m_{3/2}$ is the gravitino mass.
These contributions were first noticed in \cite{macintire}, but
were fully appreciated in the work of
\cite{randallsundrum,lutyetal}.
Further theoretical insight into the anomaly has been provided by
the work of \cite{baggeretal,gaillardnelson}.
These authors provided a thorough
understanding of the nature of the anomaly, and also gave certain
conditions under which the one- and two-loop formulas
(\ref{anomalymasses})
%\cite{randallsundrum,lutyetal}
are applicable.

There are, however, several questions which one might raise about
the sequestered argument applied to BWSB.
Even with the separable form (\ref{fsep}) the fields
on the visible and hidden sector branes
are coupled through current--current interactions coming from the
third term in the Lagrangian (\ref{cremmerlag}).
So bulk locality does not forbid brane--brane interactions
which appear non-local from the microscopic point of view.
This is not surprising
since there is no sense in which the branes are far apart
in the low energy four--dimensional theory. From the microscopic
point of view these brane--brane couplings
must arise from exchange of bulk fields between the branes.

To explore these questions, the only available framework is string
or M theory.  We will see that
the specific form of the brane--brane couplings
depend on what fields are present in the bulk
and how these couple to brane fields.
In addition, with a co-dimension one bulk such as in heterotic
Horava-Witten theory, the brane--brane couplings induced
by exchange of bulk fields might be expected to grow, or at least
remain constant with brane separation.
With higher co-dimension, brane--brane couplings would be expected
to be suppressed by the internal geometric volume.
Indeed, we will show in the next section that this is the case.
However, the four--dimensional gravitino mass is also suppressed
by the same power of the internal volume.
Soft masses arising from such brane--brane interactions are then not
necessarily suppressed with respect to the gravitino mass.

%The specific form of the supergravity $f$ function, or equivalently
%Einstein frame Kahler potential, then depends on the specific
%BWSB background, as illustrated in the following sections.
%The use of string and M-theory backgrounds provides a consistent
%description of the ultraviolet physics.
%This has the advantage microscopically that the form of the
%coupling of bulk and brane fields
%is specified in such backgrounds, and need not be guessed in
%an effective operator analysis within a low energy
%supergravity theory.

% ------------------------------------------------------------

\section{Lowest Order Structure of the Kahler Potential}

A natural arena for realizing BWSB in a string theory
is with D-branes.
Another is with end of the world branes such as arise
in Horava-Witten theory obtained by an orbifold projection
of M-theory.
In this section we consider BWSB backgrounds with 16 supersymmetries.
While obviously not realistic,
these examples are instructive as they illustrate
that the sequestered intuition breaks down even in highly
supersymmetric situations, and thus cannot be robust.
In addition, in more realistic models with only 4 supersymmetries
which are obtained by projections of these models, the form of the lowest
order tree-level Kahler potential for the states which survive
the projection is inherited from underlying Kahler potential.
Additional corrections which arise in backgrounds with less
than 16 supersymmetries are discussed in the next section.

Consider, first, the Horava-Witten compactification of
M-theory on an $S^1/Z_2$ interval \cite{hv}.
This theory has two $E_8$ gauge multiplets which reside
on end of the world branes which bound the interval.
These end of the world branes may be identified with the
visible and hidden sectors.
Compactification of this theory on $T^6$ gives a four--dimensional
theory with 16 supersymmetries which completely
fixes the form of the Kahler potential.
Since the form is independent of the coupling
it is identical to the weakly coupled heterotic string theory
result.
In four--dimensional $N=1$ notation the complex moduli include
chiral
fields $T_{i \bar j}$ and $T_{ij}$ 
%T_{\bar i \bar j}$,
where $i,j=1,2,3$ are the $T^6$
complex coordinates, the dilaton $S$,
and the visible and hidden sector brane
chiral matter arising from compactification of the gauge multiplets
denoted by $Q_i$ and $\Sigma_i$ respectively.
In terms of these
\cite{polchinski},
%As in \cite{polchinski}, it is a simple matter to generalize this
%to include the fields which are invariant only under the overall
%$Z_3$ and the full $E_8$ gauge groups:
\beq
K= -\ln \hbox{det}(T_{i \bar j} + T^{\dagger}_{i \bar j} -
\hbox{tr} Q_i Q_{\bar j}^{\dagger} -
\hbox{tr} \Sigma_i \Sigma_{\bar j}^{\dagger}) -\ln(S + S^{\dagger} )
\label{tdeteqn}
\eeq
where the two traces are over $E_8$ and $E_8^{\prime}$
gauge groups respectively and
the dependence on the $T_{ij}$ moduli is suppressed. Note that
this result is explicitly invariant under the $SU(3)\times U(1)_R$
subgroup of the $SU(4)$ R-symmetry as expected for the low energy
action at the level of two derivatives.
The supergravity $f$ function associated with the Kahler
potential (\ref{tdeteqn}) is
\beq
f = -3 \left[ (S + S^{\dagger})
 \hbox{det}  (T_{i \bar j} + T^{\dagger}_{i \bar j} -
 \hbox{tr} Q_i Q_{\bar j}^{\dagger} -
\hbox{tr} \Sigma_i \Sigma_{\bar j}^{\dagger})
 \right]^{1/3}  ~.
 \label{nfourf}
\eq
The Kahler potential (\ref{tdeteqn}) is not of the no-scale
sequestered form and the $f$ function (\ref{nfourf}) is clearly not
separable.
This is true even ignoring the dilaton.
So we see that in this highly symmetric brane world model
the sequestered intuition breaks down even without the inclusion
of corrections which would generically be
present in more realistic models.

The breakdown of the sequestered intuition can be seen
directly in this example by first considering the
ten--dimensional effective action which results at
length scales long compared with the $S^1/Z_2$ interval.
This limit is relevant if the $T^6$ is much larger than the
$S^1/Z_2$ interval.
The ten--dimensional action
%What went wrong with the intuition suggesting that the
%supergravity function, $f$, should be separable?  One can identify
%several problems.  First, allowing $S$ and the other moduli to
%vary, we see immediately that what we have called the
%``supergravity frame" does not coincide with the ``sequestered frame."
%Second, the naive notion of locality already breaks down in this
%situation.  One can see this by considering the calculation of the
%effective action in more detail.  The ten dimensional effective action,
at the level
of two derivative terms involves terms which are quadratic
and quartic in the fields.
Most of these terms do not couple fields
on the different branes.
But there are Chern-Simons squared terms
which do couple gauge fields in the two $E_8$ gauge groups.
In the underlying theory these gauge fields
reside on different branes and give rise to visible and hidden sector
fields in the toroidally compactified theory.
%These terms are of the form of
%Chern-Simons squared couplings.
The existence of these terms are in fact crucial
in the derivation of the four--dimensional Kahler potential,
as explained in \cite{wittensimple}.
%For example, the Kahler potential of
%%either
%%eqns. \ref{wittenform} or
%\ref{tdeteqn} are both of infinite order in fields.
From an
eleven--dimensional perspective, these brane--brane interactions
arise because the brane Chern-Simons terms act as a source
for the bulk three-form potential \cite{ovrutten}.
The resulting constant bulk four-form field strength generates
the Chern-Simons squared couplings between the branes.
So even though the branes are physically separated,
the visible and hidden sector fields are coupled through
the exchange of a bulk field.
This coupling and its flavor dependence
may also be understood as arising from the
exchange of bulk gauge bosons in a five--dimensional
limit which is appropriate if the $T^6$ is smaller than the
$S^1/Z_2$ interval \cite{longpaper}.

In the presence of hidden sector supersymmetry breaking
the Kahler potential (\ref{tdeteqn}) gives visible sector
tree-level mass squared eigenvalues of
\beq
m_{Q_i}^2 = m_{3/2}^2 (1,1,-2) ~.
\label{nfourmass}
\eq
These masses are of order the gravitino mass and
are non-universal \cite{tachnote}.
Without additional assumptions about flavor, the
squark and slepton mass eigenstates associated
with these eigenvalues need not be aligned with quark
and lepton eigenstates.
This would generally lead to dangerous supersymmetric
contributions to low energy flavor violating processes.
The breakdown of the sequestered intuition implies that
BWSB does not in itself provide a solution to the supersymmetric
flavor problem.

A D-brane realization of BWSB which does not have a co-dimension one
bulk may be illustrated by
considering first type I string theory with gauge group $SO(32)$
compactified on $T^6$.
This theory preserves 16 supersymmetries and therefore also
has a Kahler potential of the form (\ref{tdeteqn}).
Consider a T-duality transformation on all the $T^6$ directions.
The resulting type IIB theory has (including images)
32 D3 branes and 16 O3 orientifold planes.
Separating the D3 branes into two groups provides a model
of the visible and hidden sector branes.
In the type I description this corresponds to turning on
Wilson lines.
The Kahler potential (\ref{tdeteqn}) is invariant under this
T-duality since it is a symmetry of the four--dimensional theory;
it is unaffected by Wilson lines
and is not of the sequestered form.

%Another situation where one can describe such a setup arises in
%the Type I$^\prime$ theory.  One can again consider a toroidal
%compactification of this theory to four dimensions.  The Kahler
%potential must be the same as that of the strongly coupled heterotic
%theory.  This follows from
%our remark that the Kahler potential is unique; it also
%follows from a sequence of dualities.  So again, the Kahler
%potential does not take the sequestered form.  To understand
%microscopically why this is the case, it is simplest
%to start with the Type I theory.  The low energy theory, again,
%has $N=4$ supersymmetry, and its structure can be determined by
%compactifying the 10-dimensional action.  As a result, the terms
%with up to two derivatives are identical to those we have
%encountered above.  Now, performing a T-duality transformation on all
%of the compact dimensions, we
%can obtain a setup with $D3$-branes.  Turning on Wilson lines in
%the original picture corresponds to separating the branes in the
%$T$-dual picture.  We can, in this way, obtain two separated
%branes with different gauge groups on each brane.  Again, the
%Kahler potential is not of the sequestered form.

The origin of the brane--brane interactions in this example
may be understood by considering the simpler case
of type I theory on $S^1$.
In this theory, there are, as in the heterotic case, Chern-Simons squared
terms.
These terms remain as Wilson lines are turned on.
Now consider the T-dual type I$^{\prime}$ description.
In this theory there are (including images) 32 D8 branes and
2 O8 orientifold planes.
A Wilson line in the type I description corresponds to motions
of the D8 branes in the type I$^{\prime}$ description.
In the low energy four--dimensional theory resulting from
further compactification on $T^5$, fields which reside on
separated groups of D8 branes therefore have brane--brane interactions
corresponding to the Chern-Simons squared couplings of the type I
description.
Microscopically these can be understood in the type I$^{\prime}$
description as arising from the exchange of the bulk Ramond
two-form potential in a manner analogous to the bulk exchange in
the Horava-Witten model \cite{longpaper} .
The type IIB model above is obtained from the type
I$^{\prime}$ model on $S^1$ by T-duality on the remaining
$T^5$ directions.
The original type I theory is also dual to the Horava-Witten
theory by type I--heterotic duality in the strongly coupled limit.
So all the BWSB backgrounds of this section are related by dualities.

In the type IIB model the brane--brane interaction terms are suppressed
by the internal volume \cite{longpaper}.
Since the four--dimensional gravitino mass is also suppressed
by the internal volume, hidden sector supersymmetry breaking
then translates into squark and slepton masses of order
the gravitino mass.
It is not surprising
that such volume-suppressed terms are present,
and that they violate naive notions of locality.
%It is also not too difficult
%to trace how these terms arise.
At the level of the brane--brane interaction amplitude,
these interactions arise in the open--string channel from
the quantum one--loop amplitude of massive strings which stretch between
the branes.
The volume dependence may be understood as arising
from the sum over open string winding modes \cite{kramer}.

In sum, already at the leading level, the sequestered form of the
Kahler potential does not hold in BWSB backgrounds where one can
calculate.
At the microscopic level the brane--brane interactions which
lead to the tree-level scalar masses may be understood, at least
in some descriptions, as arising from exchange of bulk fields.
One might imagine that the arguments
leading to the sequestered no-scale form of the Kahler
potential might then hold
in a theory with a very minimal set of bulk fields.
%such as a hypothetical pure five--dimesional supergravity.
With a flat interior this does in fact occur at lowest order
for a hypothetical pure five--dimensional supergravity
with end of the world branes \cite{lutysundrum}.
This may be traced to the fact that in the dimensional
reduction from five to four dimensions, the single volume
modulus turns out not to acquire a kinetic term \cite{longpaper}.
But this special situation does not arise for
reduction from higher dimensions to four dimensions.
In addition, as discussed in the next section, a finite brane
tension leads to warping of the internal geometry
which gives additional contributions to tree-level scalar masses.
%It seems likely that such effects occur in any
%brane configuration.
Given that in the would-be five-dimensional model
there is no modulus on which the brane tension could depend,
and therefore be parametrically small, it would be surprising, again,
to find a sequestered no-scale Kahler potential in the full theory
unless the brane
tensions happened to vanish for some reason.
%A more detailed study of this hypothetical five dimensional
%case appears in \cite{longpaper}.

% --------------------------------------------------------

\section{Beyond the Leading Order}

The BWSB backgrounds of the previous section
preserve 16 supersymmetries for which the
form of the Kahler potential is determined completely by
supersymmetry.
In theories with less supersymmetry,
the leading form of the tree-level Kahler potential is often
inherited from the form dictated by the extended supersymmetry
of some underlying theory.
However, with less supersymmetry the Kahler potential is not
protected, and corrections to the leading tree-level results
should be expected.
There are at least two
situations where such corrections have been analyzed:  configurations
of branes in Type II theory with $8$ supersymmetries \cite{brodie}, and
Horava-Witten theory compactified on Calabi-Yau spaces with
4 supersymmetries \cite{ovrutetal}.
We review these and discuss their implications here.
In both cases,
we will see that these corrections do not take the sequestered
form.
The corrections may be understood as due to the distortion (warping) of
bulk background fields and geometry by brane sources.  This warping leads to
modifications of the Kahler potential, which need not be
universal, much less of the sequestered form.
%In this paper we will content ourselves with a brief summary
%of the main results with details presented elsewhere \cite{longpaper}.

Consider, first, a Type II D-brane configuration
with a source hidden sector Dp$^\prime$ brane and
a probe visible sector Dp brane \cite{brodie}.
The metric line element and dilaton backgrounds of the source
Dp$^{\prime}$-brane at distances large compared to the string
scale are
\beq
ds^2 = f(r)^{-{1 \over 2}} dx_{\|}^2 + f(r)^{1 \over 2} dx_{\perp}^2
~~~~~~e^{-2\phi} = f(r)^{p^\prime -3 \over 2}
\eeq
with
\beq
f(r) = 1 + g_s \left(
  {\sqrt{\alpha^\prime} \over r} \right)^{7-p^{\prime}} ~.
\eeq
On the visible sector probe Dp-brane world volume, these background bulk fields
yield possible corrections to the
potential and visible sector kinetic terms.
Evaluating the Dp-brane Dirac-Born-Infeld action in these background
fields
\beq
S_p =- T_p\int d^{p+1}x ~e^{-\phi} \sqrt{{\rm det}~ (h_{\mu \nu}
+ {1 \over 2} F_{\mu \rho} F^{\rho} _{\nu})}
\eeq
where $h_{\mu \nu}$ is the induced metric,
yields
\beq
S_p = f(r)^{p^\prime -3 \over 4} f(r)^{-({p+1 \over 4})}
 \left[ 1+f(r)
\left({1 \over 2} \partial_{\mu} X^i \partial^{\mu} X_i - {1 \over 4}
F_{\mu \nu} F^{\mu \nu}
\right)+ \cdots \right]
\label{dbranecorr}
\eeq
with indices now raised and lowered using the Minkowksi metric.
For $p=p^\prime$, the D-branes
%are in a ${1 \over 2}$ BPS configuration and
preserve 16 supersymmetries.
In this case from (\ref{dbranecorr}) it is apparent that the
Dp-brane world volume kinetic terms receive no corrections.
The Kahler metric is flat and the Kahler potential is exact as
required with 16 supersymmetries.
(The correction to the potential term in (\ref{dbranecorr})
is canceled by the exchange of
the RR p-form antisymmetric tensor field for $p=p^\prime$).
For $p=p^\prime-4$, the configuration
%is ${1 \over 4}$ BPS and
preserves 8 supersymmetries.
In this case, the dilaton contribution cancels the gravitational
contribution to the potential, but there is a correction to the
Dp-brane world volume kinetic terms.
The flat inherited Kahler metric
for an isolated Dp-brane, which alone would preserve 16 supersymmetries,
is modified by the background fields generated
by the Dp$^{\prime}$-brane.
The inherited Kahler potential is therefore modified by brane--brane
interactions in the configuration with only 8 supersymmetries.
This modification is due to the distortion of space caused by the
source brane at the position of the probe.
%In the world--volume
%theory this appears in the dependence of the world volume
%couplings
%on the hidden sector moduli .
This effect appears to be general.

Additional corrections to the Kahler potential could arise
in backgrounds which preserve only 4 supersymmetries, such
as the
Horava-Witten theory, compactified on a Calabi-Yau
space \cite{ovrutetal}.
%In this paper we will content ourselves with a brief summary
%of the main results with details presented elsewhere \cite{longpaper}.
%Here we will only review the main results
%of that paper.  A more detailed discussion appears in
%\cite{longpaper}.
Here we will content ourselves with a brief summary
of the main results, leaving the details for \cite{longpaper}.
In this theory, as explained in
\cite{wittency}, classical solutions for the bulk fields may be
obtained by systematically expanding
in powers of $\epsilon = T/S$.
To zeroth
order, the solution is
a direct product of the metric of the Calabi-Yau space, with
gauge fields, say, equal to the spin connection on one of the
walls, and a flat eleventh--dimension.
At next order, the space is distorted by
the presence of non-zero tension of the walls, and is a general
fibration of a Calabi-Yau space over the M-theory interval.
It is important that the shape of the Calabi-Yau
is modified along the interval in general.
This distortion of the metric
leads to modifications of the
Kahler potential. In particular, the kinetic
terms for fields localized on the walls receive
corrections that depend on the distortion of the Calabi-Yau.
%Because the shape of the Calabi--Yau changes
%across the M--theory interval these modifications depend on
%the expectation value of the Kahler forms.
Because the zero mode
wave functions on the Calabi-Yau manifold are not uniform,
and in the absence of a flavor symmetry not identical,
%As a result, for example, the
%the corrections to the zero mode kinetic terms,
the corrections to the zero mode kinetic terms
are not in any sense universal.
In the picture suggested by \cite{wittency}, the parameter $\epsilon$ is
not terribly small (of order $1/3$), and these corrections are
likely to be substantial \cite{banksdine}.  Thus, warping of space
leads to large, non-universal corrections to the tree-level
scalar masses.

It is worth noting that these remarks regarding non-universality
are also relevant to another
proposal for understanding degeneracy of squarks and sleptons:
dilaton dominated supersymmetry breaking.
In the
heterotic string at weak coupling, it has long been known that
if the dilaton F-term is
the principle source of supersymmetry breaking, squark and slepton
masses are universal at tree-level \cite{dilatondomination}.
Indeed, this is the only proposal
which realizes, in a fundamental theory,
what has traditionally been called
gravity mediation.  The scenario, if realized, is quite predictive.
The question has always been:  given that one does not expect the
string coupling to be weak, how large are the corrections to this
picture likely to be.
An optimistic assumption based on the weakly coupled
picture has been that
these corrections would be of order $\alpha_{\rm GUT} / \pi$.
This would provide just enough degeneracy to avoid
dangerous sflavor violation \cite{louisnir}. The
analysis above of the strongly coupled Horava-Witten limit
suggests however, that the
corrections could, in practice, be much larger for the actual
value of $T/S$.

% ---------------------------------------------------------

\section{Conclusions}

There are many string and M-theory backgrounds which can provide
models for BWSB, and in which
the Kahler potential can be calculated in a systematic
fashion.
None of the examples presented here, nor ones with similar
properties, yield Kahler potentials of the sequestered
form.
Non-universal tree-level squark and slepton masses generally
arises from BWSB.
In each case, it is possible to understand the microscopic
origin of the brane--brane interactions which lead to these
masses in the presence of hidden sector supersymmetry breaking.
It is not surprising that the sequestered intuition generally
breaks down since in the low energy four--dimensional theory
there is no sense in which the visible and hidden sector
branes are separated.

The effects discussed here can generally be understood as
arising from exchange of bulk supergravity fields.
This might lead one to speculate that perhaps the sequestered
form would hold in a theory with a minimal number of bulk fields.
In particular, one could conceive of a
background which reduces to pure five--dimensional supergravity
with $8$
supersymmetries in a five--dimensional bulk, broken to four
dimensions by end of world branes.
If the fifth dimension
is flat and there is only a single
overall volume modulus, $T$,
the no-scale sequestered form is in fact obtained at the classical level
\cite{lutysundrum}.
This form might also be obtained more generally in BWSB models
with background fluxes which stabilize all moduli but
a single overall volume modulus \cite{fluxmodels}.
The discussion of the previous section, however, suggests that in
any such situation, the branes will generally
warp the internal geometry.  This in
turn will induce $T$-dependence in the kinetic terms and couplings
of the fields on the brane
%and the sequestered form will not
%hold,
and tree-level masses will result.
It is difficult to test these ideas, at least in the case of pure
five--dimensional supergravity
since at present there are no known string or M-theory backgrounds
of this type, and certainly no realization within a controlled
approximation \cite{harvey}.
%In \cite{longpaper} we will explore a number of
%models for this phenomenon, and, while we will not be able to rule
%out the possibility of a sequestered Kahler potential, find that
%it is unlikely.

%It is amusing to note that
%Witten long ago
%discussed a toy model for heterotic string
%compactification\cite{wittensimple},
%which
%leads (with slight modification) to a Kahler potential of the form
%\beq
%K = - 3 \ln(T+ T^\dagger - Q^\dagger Q - \Sigma^\dagger \Sigma)
%\label{wittenform}
%\eeq
%which is of the sequestered form.  It should be noted that there
%were two crucial elements in this analysis:

Broad classes of string models have matter fields localized on
separated branes, and thus
BWSB seems a plausible outcome of string theory.
%As proposed, it provides a predictive
%model for the squark and slepton masses, which potentially solves
%the supersymmetric flavor problem.  While it would be exciting to
%derive such a prediction from string theory, it is
%satisfying that we can argue that such a phenomenon is {\it not} a
%robust feature of the theory.
However, the results presented here suggest that the model building
and phenomenology of BWSB are similar to standard (super)gravity
mediation scenarios rather than anomaly mediation.
It suggests that we must look in
other directions for the solution to the supersymmetric flavor problem.

\noindent

{\bf Acknowledgements:}

\noindent

We would like to thank Z. Chacko, C. Csaki, and S. Kachru for discussions.
We would also like to thank the M--Theory Workshop held
at the ITP Santa Barbara,
and the Aspen Center for Physics, for
their hospitality while some of this work was completed.
The work of A.A., M.D., and M.G. was
supported in part by a grant from the US
Department of Energy.
The work of S.T. was supported in part by the US National
Science Foundation under grants PHY98-70115 and PHY99-07949,
the Alfred P. Sloan Foundation, and Stanford University through
the Fredrick E. Terman Fellowship.

%%%%%%%%%%%%%%%%%%%%%%%%%%%%%%
%  Bibliography
%%%%%%%%%%%%%%%%%%%%%%%%%%%%%%


\begin{thebibliography}{99}
\bibitem{tachnote}
The unacceptable tree-level tachyonic eigenvalue in (\ref{nfourmass})
arises from a ${\rm tr}~m^2=0$ sum rule which in this case
is enforced by the extended supersymmetry.
Projections to models with only 4 supersymmetries need not
satisfy this sum rule; general conditions on orbifold
projections to avoid tree-level tachyons are presented
elsewhere \cite{longpaper}.
\end{thebibliography}
\end{document}